\documentclass{PoS}

\newcommand{\ket}[1]{\left\vert#1\right\rangle}
\newcommand{\bra}[1]{\left\langle#1\right\vert}

\newcommand{\transition}[2]{$\left\vert#1\right\rangle\leftrightarrow\left\vert#2\right\rangle$}
\newcommand{\Ai}{\mathrm{Ai}}

\newcommand{\AiZero}{\mathrm{AiZero}}

\newcommand{\qb}{\textit{q}\textsc{Bounce} }

\title{A Gravity of Earth measurement with a \emph{q}\textsc{Bounce} experiment}

\ShortTitle{A Gravity of Earth measurement with \emph{q}\textsc{Bounce}}

\author{\speaker{Gunther Cronenberg}, Hanno Filter, Martin Thalhammer, Tobias Jenke and Hartmut Abele\\
        Atominstitut, Technische Universit\"at Wien\\
		Stadionallee 2, 1020 Wien, Austria\\
        E-mail: \email{cronenberg@ati.ac.at},\email{hfilter@ati.ac.at},\email{mthalhammer@ati.ac.at}
        \email{tjenke@ati.ac.at},\email{abele@ati.ac.at}}

\author{Peter Geltenbort\\
		Institut Laue Langevin\\
		71 avenue des Martyrs - CS 20156 - 38042 Grenoble Cedex 9 - France\\
        E-mail: \email{geltenbort@ill.fr}}

\abstract{We report a measurement of the local acceleration $g$ with ultracold neutrons based on quantum states in the gravity potential of the Earth. The new method uses resonant transitions between the states \transition{1}{3} and for the first time between \transition{1}{4}.  The measurements demonstrate that Newton's Inverse Square Law of Gravity is understood at micron distances at an energy level of $10^{-14}$ eV with $\frac{\Delta g}{g}=4\times10^{-3}$. The results provide constraints on any possible gravity-like interaction at a micrometer interaction range. In particular, a dark energy candidate, the chameleon field is restricted to $\beta<6.9\times10^{6}$ for $n=2$ (95\% C.L.). }

\FullConference{The European Physical Society Conference on High Energy Physics\\
		22--29 July 2015\\
		Vienna, Austria}

\begin{document}

\section{Introduction}

The novel and relatively recent technique of Gravity-Resonance-Spectroscopy (GRS) was demonstrated in 2011~\cite{Jenke2011}. It has numerous applications due to its intrinsic sensitivity in the micron range and has already contributed to tests on the origin of both the dark energy and dark matter content by imposing limits on axion and chameleon fields \cite{Jenke2014}. A review can be found in \cite{Abele2008}. The system is based on ultracold neutrons (UCNs), which have a velocity of around 8~m/s. These UCNs, insensitive to electric fields and shielded from magnetic gradient fields are subject to the gravitational field of the Earth. When stored above a neutron mirror, made out of a material from which the UCNs are totally reflected, the UCNs form bound, discrete energy eigen-states, see fig.~\ref{fig:states}.  Such quantum-states were demonstrated experimentally first at the Institut Laue Langevin~\cite{Nesvizhevsky2002,Nesvizhevsky2005, Westphal2007}. The $n$-th eigen state takes the form:
\begin{equation}
\psi_n(z)=c_n\Ai(\frac{z}{z_0}-\frac{E_n}{E_0}),
\end{equation}
with the characteristic length and energy scale $z_0$ and $E_0$ and a constant $c_n$. The energy of the $n$-th state is in the pico-eV range and is given by:
\begin{equation}
E_n=-\AiZero(n)\sqrt[3]{\frac{\hbar^2 m_N g^2}{2}},
\end{equation}
with the $n$-th root of the Airy function, Planck's constant $\hbar$, the neutron's mass $m_N$ and the local acceleration $g$.
The dynamics of such a wave function falling down a step was studied in~\cite{Jenke2009,Abele2009}.

Gravity-Resonance-Spectroscopy, drives transitions between the states by applying resonant oscillations.
It has now been used to serve as a pure local acceleration measurement. The transition frequency of any two gravitational bound states $i$ and $j$ is $f_{ij}=(E_i-E_j)/h$. In contrast to previous realizations, with the here presented setup the frequency depends solely on the mass of the neutron, Planck's constant and the local acceleration $g$. The setup serves as purely quantum mechanical measurement.
The experimental data shows that Newton's Inverse Square Law is verified to $\frac{\Delta g}{g}=4\times10^{-3}$ in the micron range.

\section{Method}
\begin{figure}[t]
	\begin{center}
		\includegraphics[width=.26\textwidth]{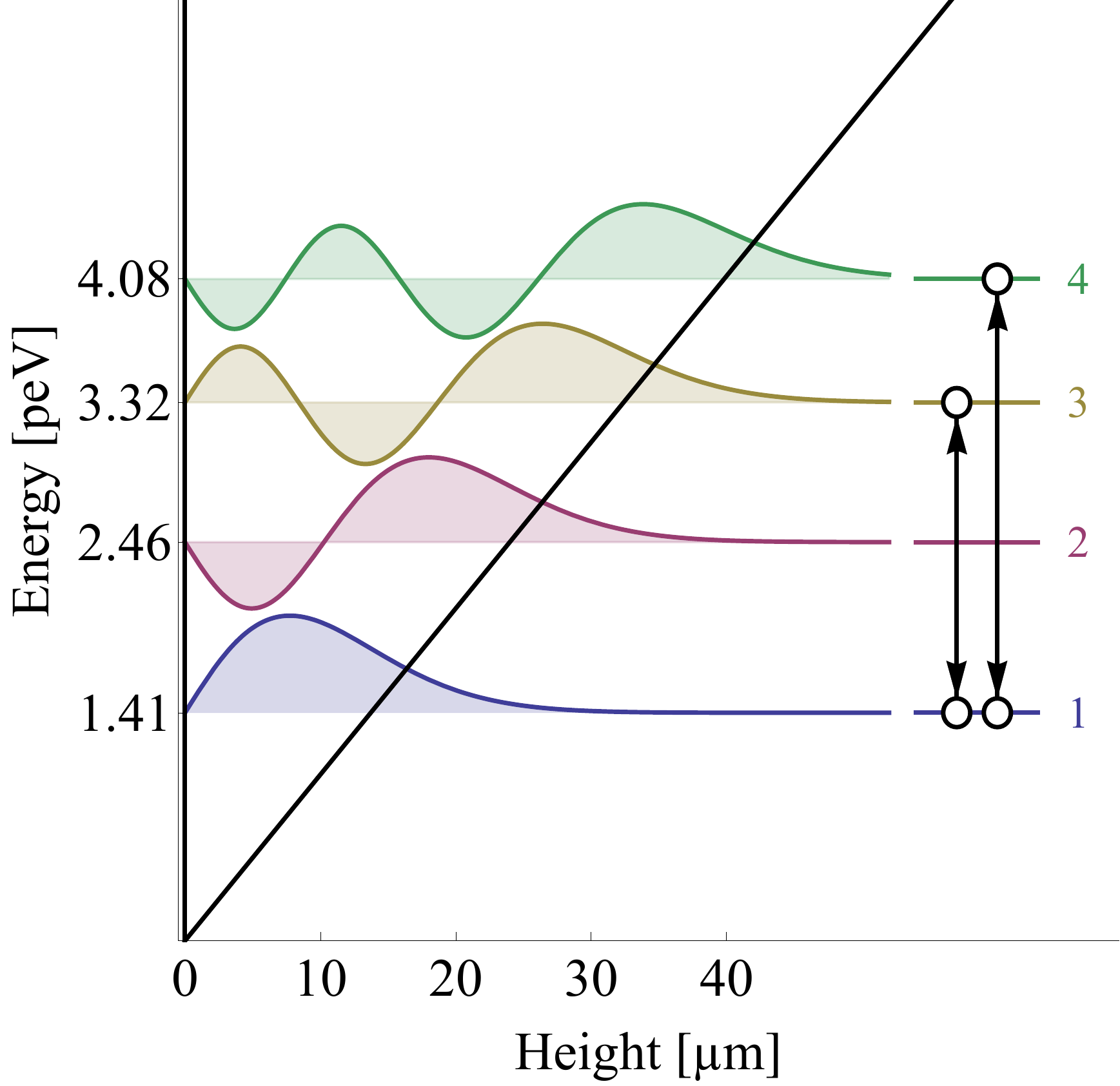}
		\includegraphics[width=.4\textwidth]{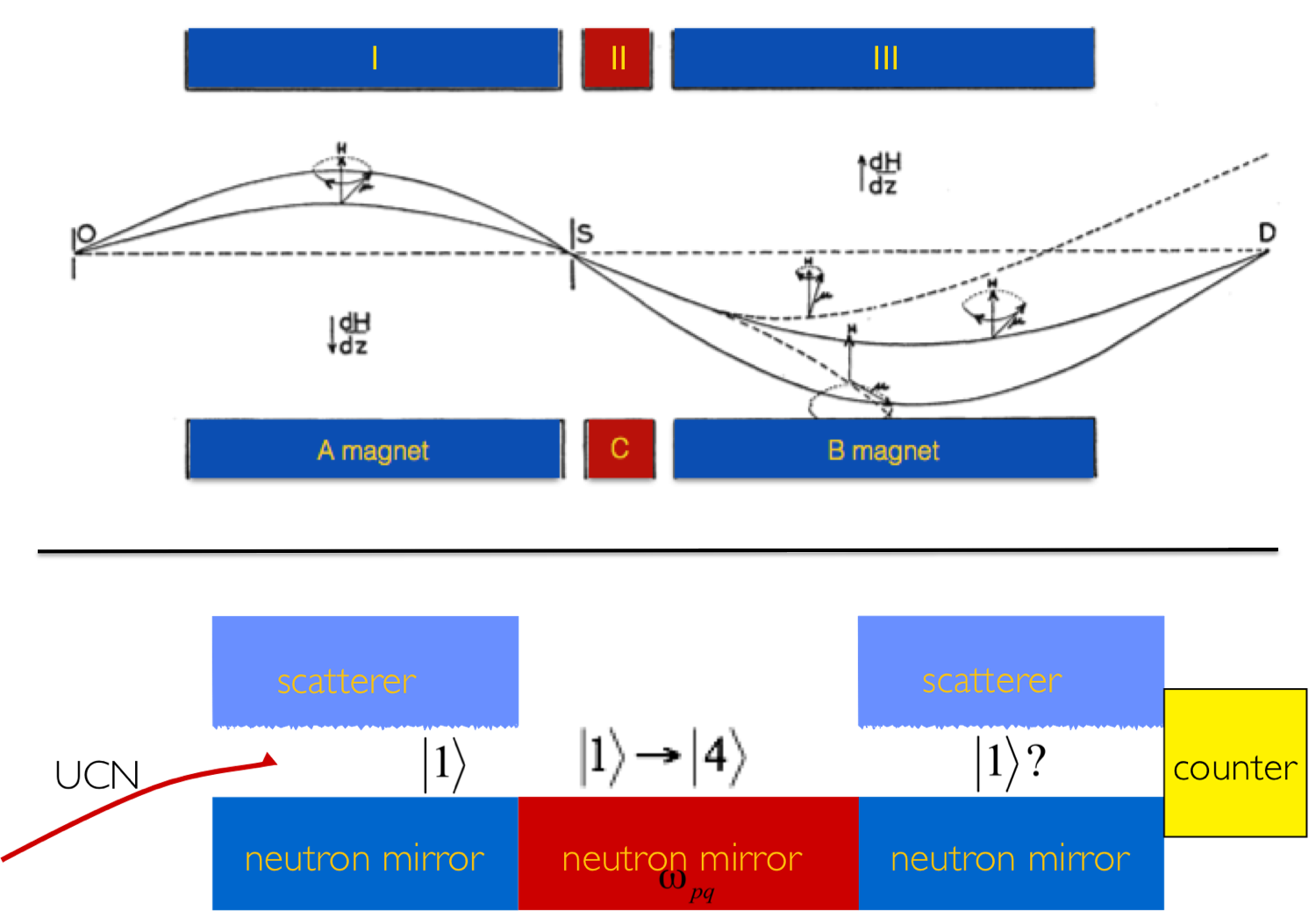}
		\includegraphics[width=.26\textwidth]{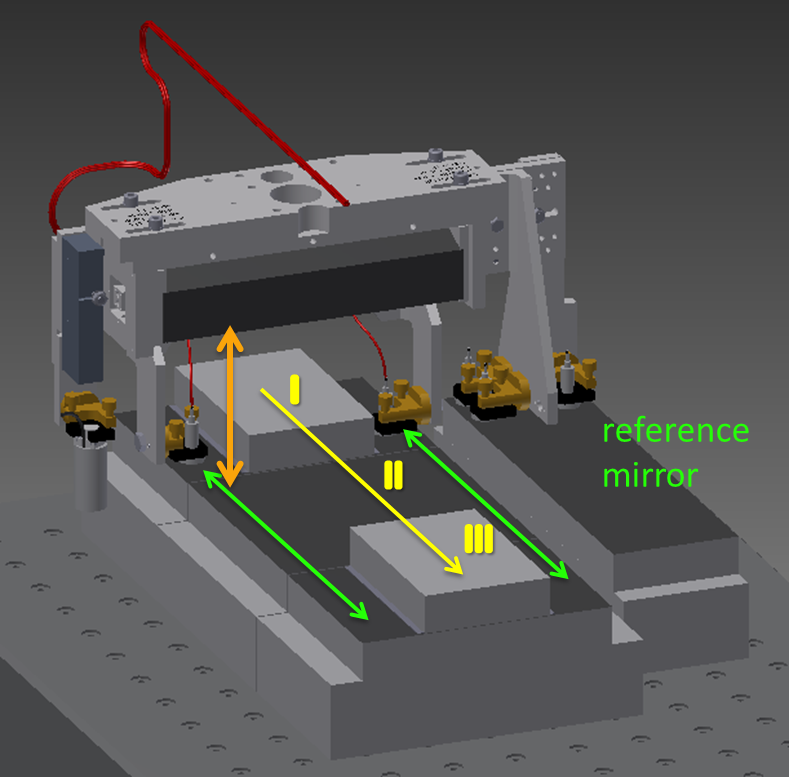}
	\end{center}
		\caption{\emph{Left:} The first five eigen states and the corresponding eigen energies of an ultracold neutron in the gravity field above a neutron mirror placed below $z=0$ which is shown as an infinite potential. The transitions \transition{1}{3} and \transition{1}{4} were successful driven upon resonance. The black lines show the potential. \emph{Middle:} A comparison of the original Rabi scheme (\emph{top}, picture adapted from~\cite{Rabi1939}) with the here presented implementation (\emph{bottom}). Both experiments are comprised of three distinct space regions, the state selection for state preparation and analysis are marked in blue and the excitation region for state transition is marked in red. \emph{Right:} The experimental setup with the three regions.}
	\label{fig:states}
\end{figure}

Here, we present a setup consisting of three regions in analogy to Rabi's original Method of Measuring Nuclear Magnetic Moment~\cite{Rabi1939} , see fig.~\ref{fig:states}. The most important difference however, is that the new setup does not use the Zeeman splitting of a spin system in a magnetic field but gravity pseudo-spin eigenstates of a neutron. Region I, formed by two stacked mirrors of 15 cm length and separated by 28 microns, was used to prepare the neutrons in the ground state. In region II only a bottom mirror with increased length of 20 cm was used which was mounted on a piezo actuator system. The position of this mirror was periodically modified in direction of the acceleration of the Earth to drive transitions from the neutron's ground state to the desired excited state. The on-following region III was identical to region I and acted as a state selector for the ground state. After the setup a neutron counter based on a Boron-10 neutron converter layer detected surviving neutrons and the events were recorded online. Previous realizations of GRS employed a one-region setup consisting of two neutron mirrors separated by roughly 30 microns. Now, for the first time, the \qb setup is realized with three distinct space regions. A significant improvement of this setup lies in the fact that no upper mirror is needed in region II compared to our previous realizations, which lead to an energy shift of the eigen-states. The previous setup added a dampening mechanism to the transition, which is avoided here without an upper mirror in the interaction region.

The oscillation behavior of the mirror in region II was monitored with a three-axes laser interferometer. Possible steps between the regions were monitored with capacitive sensors moving above the mirrors and corrected by nano-positioning tables below, see fig.~\ref{fig:states}. Step heights larger than 0.5 microns could be avoided. The experiment was placed in a vacuum of $2\times10^{-4}$ mbar. 
Furthermore, with this setup, each transition can be treated as a pure two-level system as the transition frequencies are well separated, other effects can be neglected with the current level of sensitivity~\cite{Baeßler2015a}.

The experiment took place at the ultracold neutron installation PF2 of the Institut Laue Langevin. The observed flux of ultracold neutrons through the set-up was rather low and in the order of 10 counts/1000 seconds due to the prolonged travel distance. A background optimized neutron detector was needed: both the construction and the electronics were designed with this goal in mind and the employed detector had a background rate of $r_{BG}=(641\pm17)10^{-6} \mathrm{s}^{-1}$. For a detailed description of the used detection techniques see~\cite{Jenke2013}. For the measurement scheme, the rate of the neutron flux was normalized to the rate when no oscillation was applied.

\section{Gravity Resonance Spectroscopy}
\begin{figure}[t]
	\begin{center}
		\includegraphics[width=.8\textwidth]{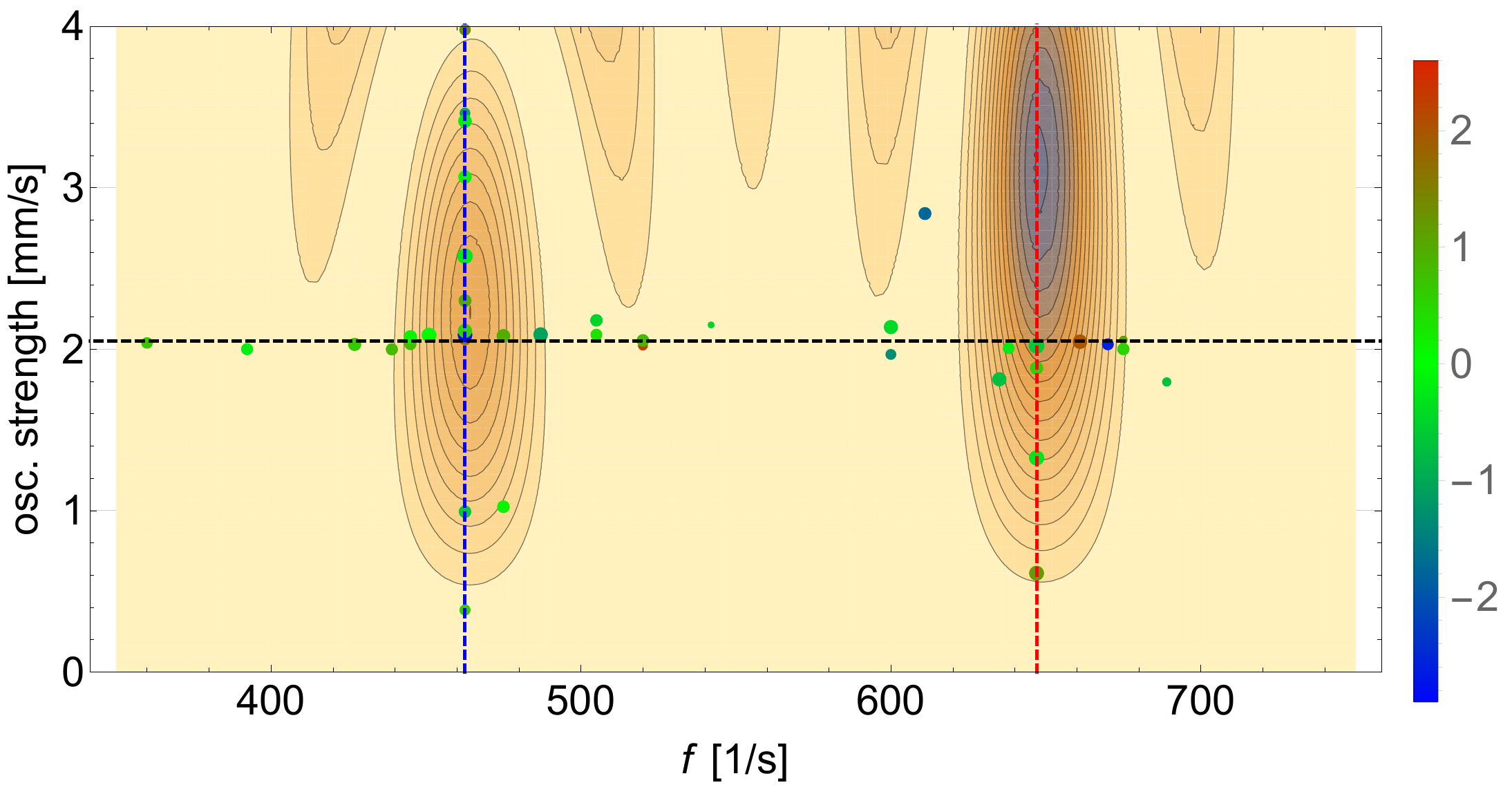}
		\includegraphics[width=.49\textwidth]{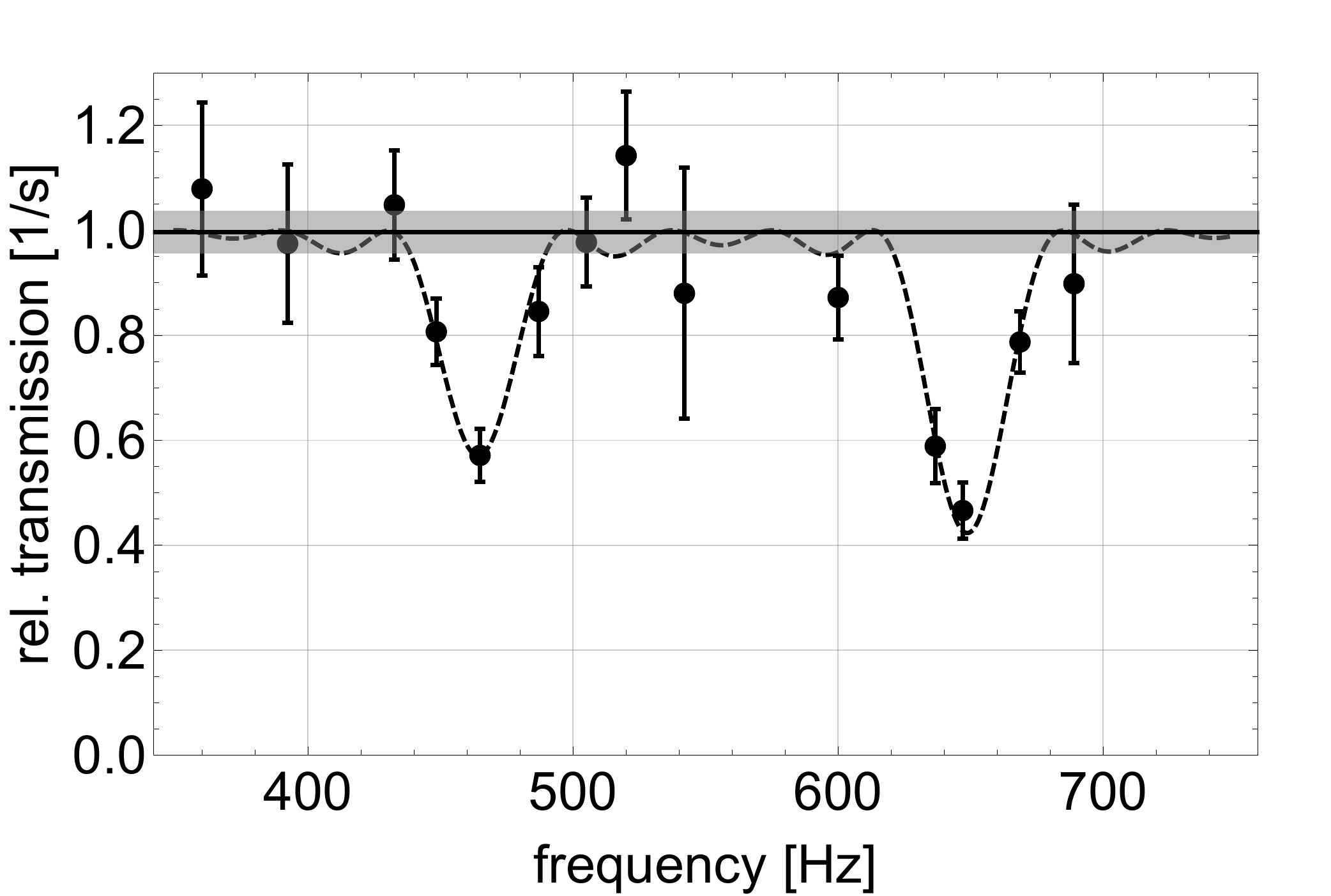}
		\includegraphics[width=.49\textwidth]{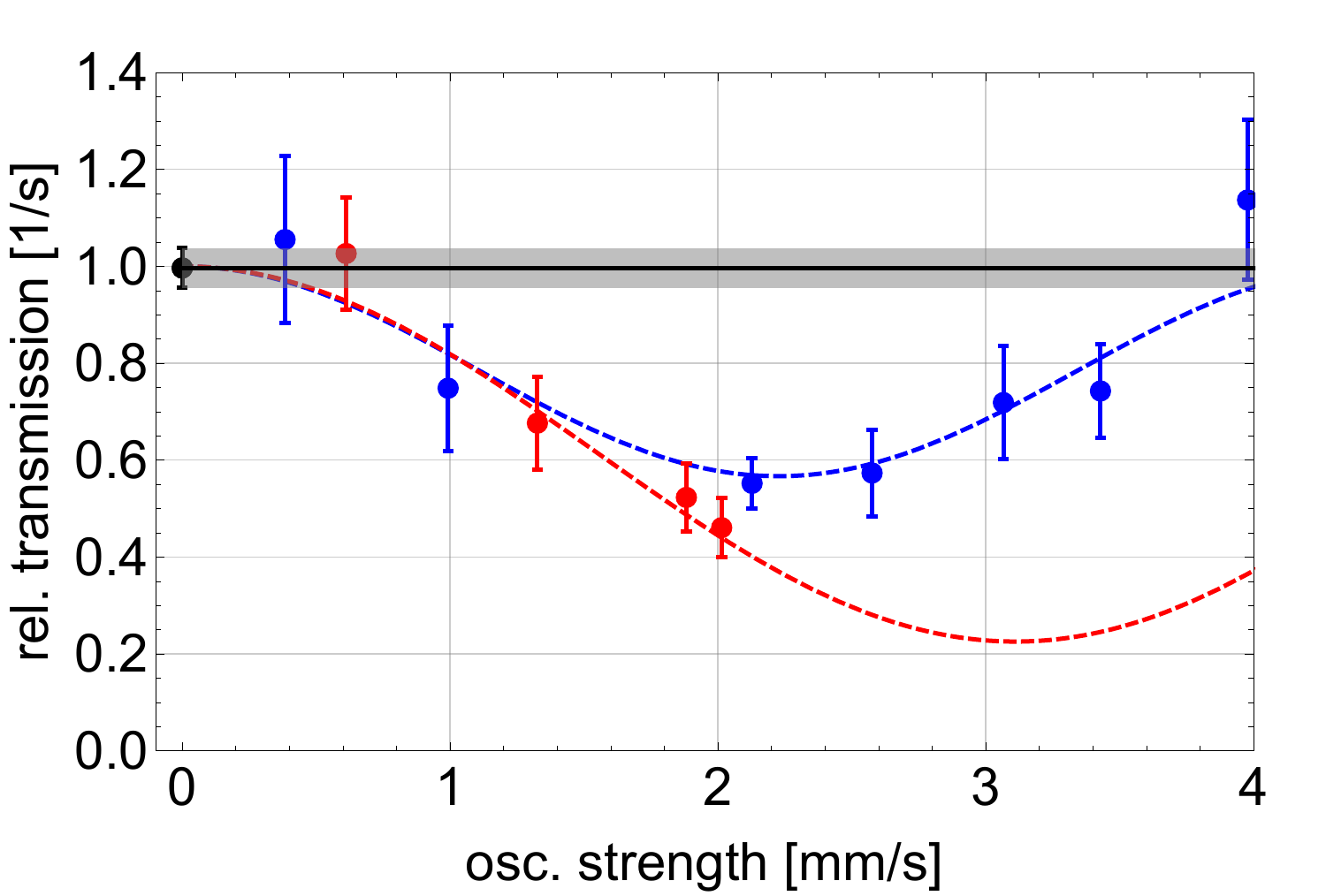}
		\caption{\emph{Top:}
			The contour plot shows the observed transmission rate in dependence of the oscillation strength and amplitude. The plot is normalised to the rate observed without any applied oscillation. The contour surface is the theoretical curve for transitions \transition{1}{3} and \transition{1}{4}with the parameters obtained by fit. Each dot represents a measurement, its size is inversely proportional to its statistical uncertainty. The colour corresponds to its normalized error with respect to the fitted curve. \emph{Lower left:} The transmission rate in dependence on the oscillation frequency for an oscillation strength of 2.05 mm/s. The curve is a cut at this strength through the contour plot. Only measurement points within a close oscillation strength are plotted. The data are binned equidistantly with 20 Hz spacing. \emph{Lower right:} The rate in dependence on the oscillation strength for the frequencies 463Hz and 647Hz which correspond to the \transition{1}{3} and \transition{1}{4} transitions. The curves are cuts of constant frequencies in the contour plot. All shown error bars indicate the statistical 1-$\sigma$ error.
			}
		\label{fig:resultate}
	\end{center}
\end{figure}

In several measurements we modified both the frequency $f$ and the oscillation strength $A$ of the applied mechanical oscillation and recorded the observed neutron rate (see fig.~\ref{fig:resultate}). The incoming neutron flux was accounted for with a beam monitor and corrected for the observed background. At the frequencies $f_{13}= 464.1\pm 1.2$~Hz and $f_{14}= 648.8\pm 1.6$~Hz significant drops in the count rate were observed which correspond to the transitions \transition{1}{3} and \transition{1}{4}, the latter one being observed for the first time. With increasing oscillation strength, the count rate drops for both frequencies until a $\pi$-flip is induced. This takes place at $a_{\pi} = $2.1~mm/s for the \transition{1}{3} transition. For the first time, the initial state could be restored with a $2\pi$ flip for the \transition{1}{3} transition.
The data was fitted simultaneously and yielded an energy scale of $E_0 = 0.60$~peV, which is connected with the energy difference of the involved states:

\begin{equation}
\Delta E_{ij}=E_0 \left( \Ai_0(i)- \Ai_0(j)\right),
\end{equation}

with $\Ai_0(i)$ being the $i$-th zero point of the Airy function. 
The local acceleration obtained is $g = 9.84\pm0.04 \mathrm{m/s^2}$ (1-$\sigma$ statistical error, the systematic error is smaller than that). A plot of the experimental data together with the obtained fit as a function of the applied oscillation strength and frequency can be seen in fig.~\ref{fig:resultate}.
The data is compatible with Newton's Inverse Square Law and demonstrates its applicability at a level of $4\times10^{-3}$.

\section{Limits on fifth forces}
At low energies the effects of new physics can be described by an effective field theory. Typically, to leading order the corrections caused by new physics are Yukawa-like. It has been shown that most deviations from Newton's inverse square law can be parametrized generically in the form of a Yukawa-term added to the potential:

\begin{equation}
V(z)=-\frac{GMm}{r}\left(1+\alpha e^{-z/\lambda}\right),
\end{equation}

with strength $\alpha$ and interaction range $\lambda$. Such deviations can arise for example due to large extra dimensions or any yet unknown model. The additional force would be sourced by the bottom mirror in region II and leads to an additional potential seen by the neutron. As a result the energy of the states would be shifted in a characteristic way. This would be detectable as an energy shift of the transition \transition{i}{j} that gives to first order perturbation theory:

\begin{equation}
\Delta E_{ij}=-2\pi \alpha G \rho m_N \lambda^2\left( \bra{\psi_j} e^{-R/\lambda}\ket{\psi_j} - \bra{\psi_i} e^{-R/\lambda}\ket{\psi_i}\right),
\end{equation}
where $\rho$ is the material density and $R$ the distance between neutron and the material.

\begin{figure}[t]
	\includegraphics[width=.49\textwidth]{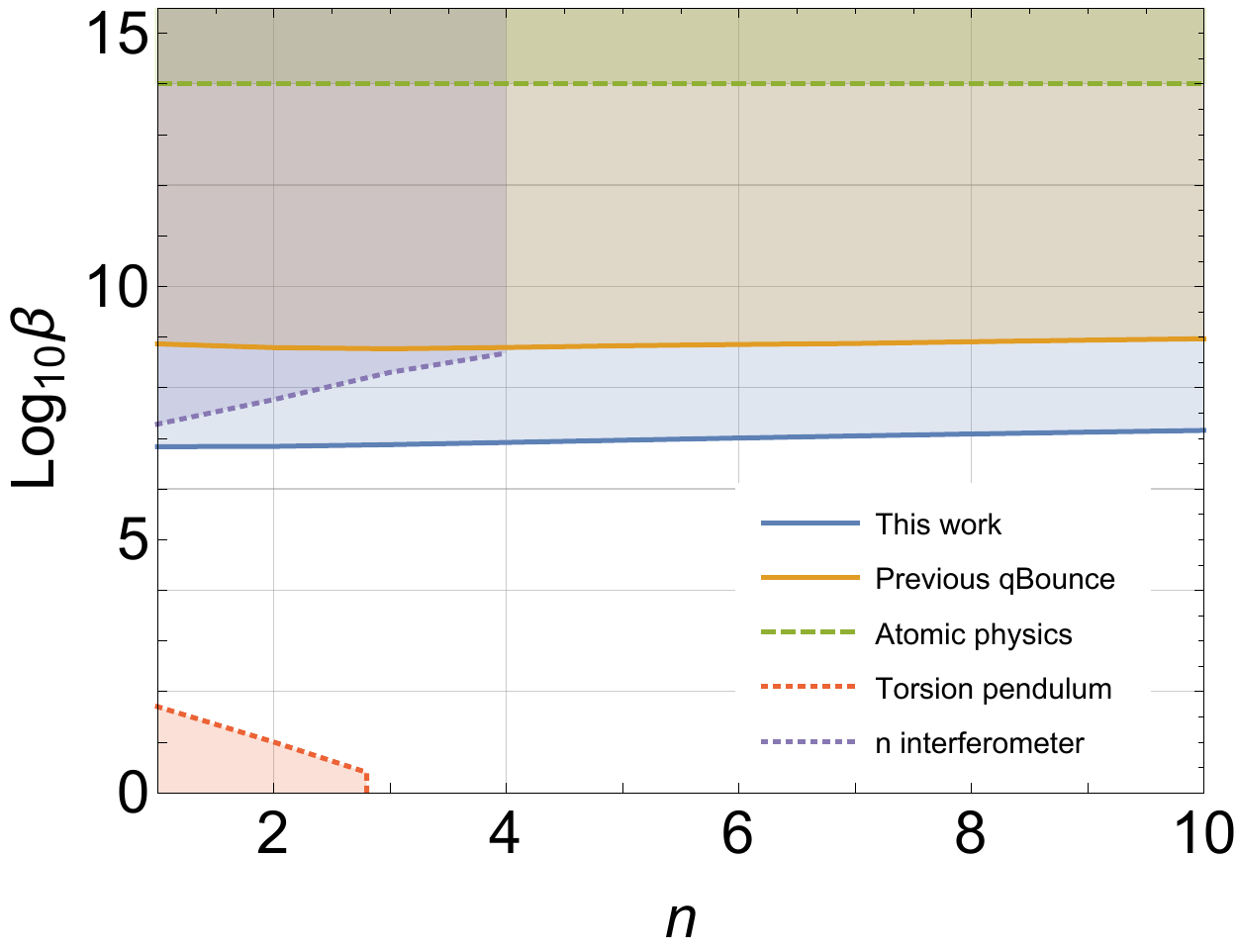}
	\includegraphics[width=.49\textwidth]{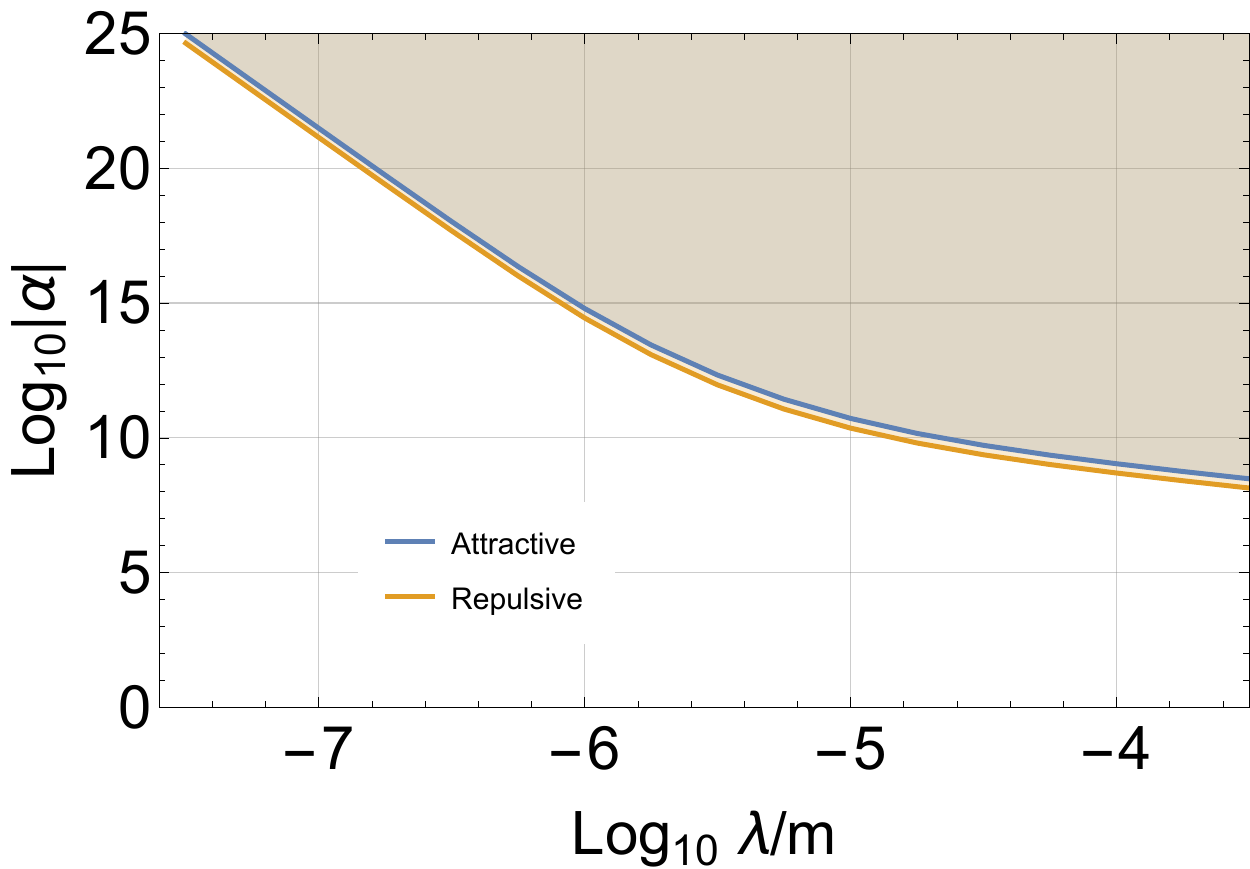}
	\caption{\emph{Left:}
		Chameleon exclusion plot for the parameter range of $\beta$ and $n$. The solid line displays the limit obtained for a confidence level of 95.45\%. The dashed line shows the limit obtained from atomic physics \cite{Brax2011}, the upper solid line the limit previously obtained with GRS in a more compact setup \cite{Jenke2014}. The purple dashed line is the limit derived with a neutron interferometer \cite{Lemmel2015}. The lower bound is obtained from torsion pendulum experiments \cite{Adelberger2009}.	\emph{Right:} The exclusion plot for Yukawa-like forces with strength $\alpha$ dependent on the range $\lambda$ at a confidence level of 95.54\%. The line for an attractive (repulsive) strength with $\alpha>0$ $(\alpha<0)$ is shown in blue (orange).}
	\label{fig:chameleon-yuk}
\end{figure}

Such a shift could be excluded, the corresponding parameter space can be found in fig.~\ref{fig:chameleon-yuk}. For a interaction length $\lambda=5.6 \mathrm{\mu m}$ a fifth force can be excluded for strengths $\alpha < -8.5\times10^{10}$ and $\alpha>2.4\times10^{11}$ with 95\% C.L.

\section{Limits on chameleon fields}
From the presented data, limits on deviation of Newton's Inverse Square Law can be imposed. One such deviation would arise due to the existence of chameleons~\cite{Khoury2004a}. The chameleon is a candidate for dark energy owing its name to the screening effect. This screening effect ensures that the chameleon field is suppressed in the vicinity of mass. While dark energy is needed at cosmological scales, this table-top experiment is sensitive to this candidate. The neutron is not being screened and thus a perfect probe. The ideal suitability of UCNs for the chameleon detection has been proposed~\cite{Brax2011} and previous limits have been already improved with GRS \cite{Jenke2014}. The mass of the chameleon field defines its range and the neutron mirror below the UCNs modifies the value of the field that is formed for a pressure of $10^{-4}$ mbar. The chameleon influences each neutron state individually thus shifting the transition frequencies in a specific way. The additional effective potential due to the chameleon has the form

\begin{equation}
V_{\mathrm{eff}}(z)=+\beta\frac{m_N}{M_{Pl}^{'}}\Lambda \hbar c \left( \frac{2+n}{\sqrt{2}}z\Lambda\right)^{\frac{2}{2+n}},
\end{equation}

where $\beta$ is the coupling strength, $n$ a model parameter (the Ratra-Peebles index), $\Lambda$ the cosmological constant and $M_{Pl}^{'}$ is the reduced Planck mass. The mass density of our neutron mirror made of borofloat-glass of type BK7 is $\rho$= 2.51 g cm$^{-3}$. For the index $n=2$, the chameleon can be excluded of $\beta = 6.9\times10^{6}$ (95\% C.L.), see fig.~\ref{fig:chameleon-yuk}.
Other limits on chameleon fields stem from neutron interferometer measurements~\cite{Lemmel2015} and atom interferometry~\cite{Hamilton2015}.

\section{Conclusion}
The GRS technique demonstrates impressive progress both at the experimental implementation as well as improved limits. The definite exclusion or confirmation of the chameleon seems not to far away.
Combining the energy scale measurements together with determining the length scale will allow to test the Universality of the Free Fall~\cite{Kajari2010}. This experiment shows that the realisation of a Ramsey-like spectrometer, which features additional space regions, is feasible, as proposed in~\cite{Abele2010}. Such a setup could be used to study the electric neutrality of the neutron~\cite{Durstberger-Rennhofer2011}.

\acknowledgments
We gratefully acknowledge support from the Austrian Fonds zur F\"orderung der Wissenschaftlichen Forschung (FWF) under Contract No. I529-N20, No. 531-N20 and No. I862-N20 and the German Research Foundation (DFG) as part of the Priority Programme (SPP) 1491 "Precision experiments in particle and astrophysics with cold and ultracold neutrons"; we also gratefully acknowledge support from the French L'Agence nationale de la recherche ANR under contract number ANR-2011-ISO4-007-02, Programme Blanc International - SIMI4-Physique.


{\small\bibliography{library}}

\begin{thebibliography}{10}

\bibitem{Jenke2011}
T.~Jenke, P.~Geltenbort, H.~Lemmel, and H.~Abele.
\newblock \emph{{Realization of a gravity-resonance-spectroscopy technique}}.
\newblock Nature Physics, \textbf{7}~(6):468 2011.

\bibitem{Jenke2014}
T.~Jenke, G.~Cronenberg, J.~Burgd{\"{o}}rfer, L.~A. Chizhova, P.~Geltenbort,
  A.~N. Ivanov, T.~Lauer, et~al.
\newblock \emph{{Gravity Resonance Spectroscopy Constrains Dark Energy and Dark
  Matter Scenarios}}.
\newblock Physical Review Letters, \textbf{112}~(15):151105 2014.

\bibitem{Abele2008}
H.~Abele.
\newblock \emph{{The neutron. Its properties and basic interactions}}.
\newblock Progress in Particle and Nuclear Physics, \textbf{60}~(1):1 2008.

\bibitem{Nesvizhevsky2002}
V.~V. Nesvizhevsky, H.~G. B{\"{o}}rner, A.~K. Petukhov, H.~Abele, S.~Baessler,
  F.~J. Ruess, T.~St{\"{o}}ferle, et~al.
\newblock \emph{{Quantum states of neutrons in the Earth's gravitational
  field.}}
\newblock Nature, \textbf{415}~(6869):297 2002.

\bibitem{Nesvizhevsky2005}
V.~V. Nesvizhevsky, A.~K. Petukhov, H.~G. B{\"{o}}rner, T.~A. Baranova, A.~M.
  Gagarski, G.~A. Petrov, K.~V. Protasov, et~al.
\newblock \emph{{Study of the neutron quantum states in the gravity field}}.
\newblock The European Physical Journal C, \textbf{40}~(4):479 2005.

\bibitem{Westphal2007}
A.~Westphal, H.~Abele, S.~Bae{\ss}ler, V.~Nesvizhevsky, K.~Protasov, and
  A.~Voronin.
\newblock \emph{{A quantum mechanical description of the experiment on the
  observation of gravitationally bound states}}.
\newblock The European Physical Journal C, \textbf{51}~(2):367 2007.

\bibitem{Jenke2009}
T.~Jenke, D.~Stadler, H.~Abele, and P.~Geltenbort.
\newblock \emph{{Q-BOUNCE - Experiments with quantum bouncing ultracold
  neutrons}}.
\newblock Nuclear Instruments and Methods in Physics Research Section A:
  Accelerators, Spectrometers, Detectors and Associated Equipment,
  \textbf{611}~(2-3):318 2009.

\bibitem{Abele2009}
H.~Abele, T.~Jenke, D.~Stadler, and P.~Geltenbort.
\newblock \emph{{QuBounce: the dynamics of ultra-cold neutrons falling in the
  gravity potential of the Earth}}.
\newblock Nuclear Physics A, \textbf{827}~(1-4):593c 2009.

\bibitem{Rabi1939}
I.~Rabi, S.~Millman, P.~Kusch, and J.~Zacharias.
\newblock \emph{{The Molecular Beam Resonance Method for Measuring Nuclear
  Magnetic Moments. The Magnetic Moments of Li63, Li73 and F199}}.
\newblock Physical Review, \textbf{55}~(6):526 1939.

\bibitem{Baeßler2015a}
S.~Bae{\ss}ler, V.~V. Nesvizhevsky, G.~Pignol, K.~V. Protasov, D.~Rebreyend,
  E.~A. Kupriyanova, and A.~Y. Voronin.
\newblock \emph{{Frequency shifts in gravitational resonance spectroscopy}}.
\newblock Physical Review D, \textbf{91}:1 2015.

\bibitem{Jenke2013}
T.~Jenke, G.~Cronenberg, H.~Filter, P.~Geltenbort, M.~Klein, T.~Lauer,
  K.~Mitsch, et~al.
\newblock \emph{{Ultracold neutron detectors based on 10B converters used in
  the qBounce experiments}}.
\newblock Nuclear Instruments and Methods in Physics Research Section A:
  Accelerators, Spectrometers, Detectors and Associated Equipment,
  \textbf{732}:1 2013.

\bibitem{Brax2011}
P.~Brax and G.~Pignol.
\newblock \emph{{Strongly Coupled Chameleons and the Neutronic Quantum
  Bouncer}}.
\newblock Physical Review Letters, \textbf{107}~(11):111301 2011.

\bibitem{Lemmel2015}
H.~Lemmel, P.~Brax, a.N. Ivanov, T.~Jenke, G.~Pignol, M.~Pitschmann,
  T.~Potocar, et~al.
\newblock \emph{{Neutron interferometry constrains dark energy chameleon
  fields}}.
\newblock Physics Letters B, \textbf{743}:310 2015.

\bibitem{Adelberger2009}
E.~Adelberger, J.~Gundlach, B.~Heckel, S.~Hoedl, and S.~Schlamminger.
\newblock \emph{{Torsion balance experiments: A low-energy frontier of particle
  physics}}.
\newblock Progress in Particle and Nuclear Physics, \textbf{62}~(1):102 2009.

\bibitem{Khoury2004a}
J.~Khoury and A.~Weltman.
\newblock \emph{{Chameleon cosmology}}.
\newblock Physical Review D, \textbf{69}~(4):1 2004.

\bibitem{Hamilton2015}
P.~Hamilton, M.~Jaffe, P.~Haslinger, Q.~Simmons, H.~Muller, and J.~Khoury.
\newblock \emph{{Atom-interferometry constraints on dark energy}}.
\newblock Science, \textbf{349}~(6250):849 2015.

\bibitem{Kajari2010}
E.~Kajari, N.~L. Harshman, E.~M. Rasel, S.~Stenholm, G.~S{\"{u}}{\ss}mann, and
  W.~P. Schleich.
\newblock \emph{{Inertial and gravitational mass in quantum mechanics}}.
\newblock Applied Physics B, \textbf{100}~(1):43 2010.

\bibitem{Abele2010}
H.~Abele, T.~Jenke, H.~Leeb, and J.~Schmiedmayer.
\newblock \emph{{Ramsey's method of separated oscillating fields and its
  application to gravitationally induced quantum phase shifts}}.
\newblock Physical Review D, \textbf{81}~(6):065019 2010.

\bibitem{Durstberger-Rennhofer2011}
K.~Durstberger-Rennhofer, T.~Jenke, and H.~Abele.
\newblock \emph{{Probing the neutron's electric neutrality with Ramsey
  spectroscopy of gravitational quantum states of ultracold neutrons}}.
\newblock Physical Review D, \textbf{84}~(3):5 2011.

\end{thebibliography}


\begin{thebibliography}{0}
\providecommand{\natexlab}[1]{#1}
\expandafter\ifx\csname urlstyle\endcsname\relax
  \providecommand{\doi}[1]{doi:\discretionary{}{}{}#1}\else
  \providecommand{\doi}{doi:\discretionary{}{}{}\begingroup
  \urlstyle{rm}\Url}\fi

\end{thebibliography}


\end{document}